\documentclass[prd,preprint, tightenlines,superscriptaddress,showpacs,byrevtex]{revtex4-2}

\usepackage{blindtext}

\usepackage{graphicx}
\usepackage{subfigure}
\usepackage{overpic}
\usepackage{epsfig}
\usepackage{dcolumn}
\usepackage{bm}
\usepackage{color}
\usepackage{siunitx}
\usepackage{lineno}
\usepackage[figuresright]{rotating}
\usepackage{diagbox}
\usepackage[normalem]{ulem}
\newcommand\redout{\bgroup\markoverwith
{\textcolor{red}{\rule[.5ex]{2pt}{0.4pt}}}\ULon}
\usepackage{xcolor}
\usepackage{threeparttable}
\usepackage {orcidlink}
\usepackage{multirow}
\usepackage[thicklines]{cancel}

\newcommand{\mrec}{M^{\rm recoil}}
\newcommand{\yones}{\Upsilon(1S)}
\newcommand{\ytwos}{\Upsilon(2S)}

\newcommand{\yfos}{\Upsilon(4S)}

\newcommand{\yns}{\Upsilon(nS)}

\newcommand{\eff}{\varepsilon}
\newcommand{\BR}{{\cal B}}

\newcommand{\pip}{\pi^+}

\newcommand{\piz}{\pi^0}
\newcommand{\kap}{K^+}
\newcommand{\kam}{K^-}

\newcommand{\GG}{\gamma\gamma}
\newcommand{\kl}{K_{\rm L}^0}

\newcommand{\etap}{\eta^{\prime}}

\newcommand{\psp}{\psi(2S)}

\newcommand{\EE}{e^+e^-}
\newcommand{\MM}{\mu^+\mu^-}

\newcommand{\pp}{\pi^+\pi^-}

\newcommand{\ppp}{\pi^+\pi^-\pi^0}
\newcommand{\kk}{K^+K^-}

\newcommand{\ccb}{c\bar{c}}
\newcommand{\qqb}{q\bar{q}}

\newcommand{\beq}{\begin{equation}}
\newcommand{\eeq}{\end{equation}}
\newcommand{\beqar}{\begin{eqnarray}}
\newcommand{\eeqar}{\end{eqnarray}}
\newcommand{\bitm}{\begin{itemize}}
\newcommand{\benu}{\begin{enumerate}}

\newcommand{\bitmb}{\begin{itemize}}
\newcommand{\benub}{\begin{enumerate}}
\newcommand{\eitm}{\end{itemize}}

\newcommand{\cm}{\si{\centi\metre}}

\newcommand{\gevcs}{\hbox{GeV}/c^2}
\newcommand{\gevc}{\hbox{GeV}/c}
\newcommand{\gev}{\hbox{GeV}}
\newcommand{\mevcs}{\hbox{MeV}/c^2}

\newcommand{\mev}{\hbox{MeV}}

\newcommand{\KM}{K^-}
\newcommand{\ks}{K_S^0}
\newcommand{\kbar}{\bar{K}}
\newcommand{\ktb}{\bar{K}^{*}(892)^{0}}

\newcommand{\rhop}{\rho^+}

\newcommand{\dsp}{D_s^{+}}
\newcommand{\dstp}{D_s^{*+}}
\newcommand{\dsstp}{D_{s}^{(*)+}}

\newcommand{\dz}{D^{0}}
\newcommand{\adz}{\bar{D}^0}
\newcommand{\dm}{D^{-}}

\newcommand{\dza}{D^{*}(2007)^{0}}
\newcommand{\adza}{\bar{D}^*(2007)^0}
\newcommand{\dzb}{D^{*}(2010)^{-}}
\newcommand{\dzab}{D^{(*)0}}
\newcommand{\adzab}{\bar{D}^{(*)0}}
\newcommand{\dmab}{D^{(*)-}}
\newcommand{\dbar}{\bar{D}}
\newcommand{\dtbar}{\bar{D}^*}
\newcommand{\dmbar}{\bar{D}^{(*)}}

\newcommand{\dsa}{D_{s1}(2536)^-}
\newcommand{\dsb}{D_{s2}^*(2573)^-}

\newcommand{\dsj}{D_{sJ}^{-}}
\newcommand{\dsjm}{D_{sJ}^{+}}

\newcommand{\ecm}{E_{\rm c.m.} }

\newcommand{\nb}{\si{\nano\barn}}

\newcommand{\fb}{\si{\femto\barn}}

\newcommand{\infb}{\fb^{-1}}

\newcommand{\xsbn}{\sigma^{\rm Born}}

\newcommand{\mc}{\multicolumn}

\def\Journal#1#2#3#4{{#1} {\bf #2} (#4) #3}

\def\NIMA{Nucl. Instrum. Meth. A}

\def\PLB{Phys. Lett. B}

\def\PRD{Phys. Rev. D}

\def\CPC{Chin. Phys. C}
\def\EPJC{Eur. Phys. J. C}

\def\PLB{Phys. Lett. B}

\def\PR{Phys. Rep.}

\def\PTEP{Prog. Theor. Exp. Phys. }

\begin{document}


\preprint{} \preprint{ \vbox{ \hbox{   }
		\hbox{Belle Preprint 		2023-12	}
		\hbox{KEK Preprint 		2023-16	}
}}

\title{
\quad\\[1.0cm]
\boldmath
Observation of charmed strange meson pair production in $\ytwos$ decays and in $\EE$ annihilation at 
$\sqrt{s} = 10.52~\gev$}

\noaffiliation
  \author{B.~S.~Gao\,\orcidlink{0000-0002-8143-0330}} 
  \author{W.~J.~Zhu\,\orcidlink{0000-0003-2618-0436}} 
  \author{X.~L.~Wang\,\orcidlink{0000-0001-5805-1255}} 
  \author{I.~Adachi\,\orcidlink{0000-0003-2287-0173}} 
  \author{H.~Aihara\,\orcidlink{0000-0002-1907-5964}} 
  \author{D.~M.~Asner\,\orcidlink{0000-0002-1586-5790}} 
  \author{V.~Aulchenko\,\orcidlink{0000-0002-5394-4406}} 
  \author{T.~Aushev\,\orcidlink{0000-0002-6347-7055}} 
  \author{R.~Ayad\,\orcidlink{0000-0003-3466-9290}} 
  \author{V.~Babu\,\orcidlink{0000-0003-0419-6912}} 
% ...
\collaboration{The Belle Collaboration}

\date{\today}

\begin{abstract}

We observe the process $\ytwos\to \dsstp \dsj$ and continuum production $\EE\to \dsstp \dsj$ at $\sqrt{s} = 10.52
~\gev$ (and their charge conjugates) using the data samples collected by the Belle detector at KEKB, where $\dsj$ is 
$\dsa$ or $\dsb$. Both $\dsj$ states are identified through their decay into $\kbar\dmbar$. We measure the products 
of branching fractions $\BR(\ytwos \to \dsstp \dsj)\BR(\dsj\to \kbar \bar{D}^{(*)})$ and the Born cross sections 
$\xsbn(\EE \to \dsstp\dsj) \BR(\dsj\to \kbar \bar{D}^{(*)})$, and then compare the ratios $R_1 \equiv \BR(\ytwos\to 
\dsstp\dsj)/\BR(\ytwos\to\MM)$ for $\ytwos$ decays and $R_2 \equiv \xsbn(\EE\to\dsstp\dsj)/\xsbn(\EE\to \MM)$ for 
continuum production. We obtain $R_1/R_2 = 9.7\pm 2.3 \pm 1.1$, $6.8 \pm 2.1 \pm 0.8$, 
$10.2 \pm 3.3 \pm 2.5$, and $3.4 \pm 2.1 \pm 0.5$ for the $\dsp\dsa$, $\dstp\dsa$, $\dsp\dsb$, and 
$\dstp\dsb$ final states in the $\dsj\to \KM \bar{D}^{(*)0}$ modes, respectively. Therefore, the strong decay is 
expected to dominate in the $\ytwos\to\dsstp\dsj$ processes. We also measure the ratios of branching fractions 
$\BR(\dsa\to \ks \dzb)/\BR(\dsa\to \KM \dza) = 0.48 \pm 0.07 \pm 0.02$ and $\BR(\dsb \to \ks \dm)/\BR(\dsb \to \KM 
\dz) = 0.49 \pm 0.10 \pm 0.02$, which are consistent with isospin symmetry. The second ratio is the first 
measurement of this quantity. Here, the first uncertainties are statistical and the second are systematic.

\vspace{1cm}

\end{abstract}

\maketitle

\section{Introduction}

Much of the plethora of new quarkonium states observed in the last decades has been studied at electron-positron 
colliders~\cite{review_xyz}. These accelerators have collected large data samples at center of mass (c.m.) energies 
($\sqrt{s}$) corresponding to both the vector charmonium and the vector bottomonium states, which are produced 
copiously due to resonance enhancements in the cross sections. A vector quarkonium state decays either 
electromagnetically through the annihilation process into a virtual photon or into three gluons mediated by the 
strong interaction. We can thereby separate the dynamics of electromagnetic and strong charmed meson production 
through complementary measurements at energies above or below the quarkonium state, which are called 
``off-resonance." Here, only quantum electrodynamics (QED) processes contribute and thus allow 
measurements free from hadronic structure effects and quantum chromodynamics (QCD) related process 
present for heavy quarkonia.

At $\sqrt{s}$ significantly above the prodcution threshold and far from quarkonium resonances, the production rates 
of $\EE\to\qqb$ are approximately proportional to the quark charge squared, so that $\EE\to\ccb$ is about 40\% of the 
total hadronic production at $\sqrt{s} = 10.52~\gev$ ($60~\mev$ below the $\yfos$). This provides an opportunity to 
study the charmed hadrons, including charmed mesons, charmed strange mesons, and charmed baryons. However, this kind 
of study was rarely done before. This is also true for the OZI suppressed hadronic decays of the narrow $\yns$ 
states; hundreds of millions of $\yns$ events have been accumulated at Belle and 
BaBar, but such studies are scarce. The open charm content of bottomonium hadronic decays 
can be used as a tool to probe the post-$b\bar{b}$-annihilation fragentation processes~\cite{Brambilla}. 
Within the QCD approach, the charm quarks are expected to be produced in $\yns$ decay only by a process in which a 
virtual timelike gluon of large invariant mass is produced in the initial decay process, and subsequently decays into 
a pair of charmed hadrons~\cite{theoretical1}. Using a data sample of $(98.6 \pm 0.9)\times 10^6$ $\ytwos$ events, 
BaBar measures $\BR[\yones \to D^{*+}X] = (2.52 \pm 0.13 \pm 0.15)\%$~\cite{up1dd}, which is considerable in excess 
of that expected from $b\bar{b}$ annihilation into a single photon. This excess is seen to be in agreement with a 
prediction based on splitting a virtual photon~\cite{theoretical3}, but appear to be too small to accommodate an 
octet-state contribution~\cite{yjzhang}. Here and hereinafter, the first uncertainty quoted is statistical while the 
second corresponds to the total systematic uncertainty. However, no more measurement on the charm hadron in $\yns$ 
decays can be found~\cite{PDG}. It is argued that the suppression of charm production on the $\yns$ resonance is at 
least consistent with the analogous case of strangeness production on $\psi$ and $\psp$, and it would be quite 
instructive to study the topology of such events where charm is actually produced~\cite{theoretical2}. 

Here, we present searches for $\dsstp \dsj$ with the subsequent decay $\dsj\to \kbar + \dmbar$ in $\ytwos$ decays and 
in continuum $\EE$ annihilation, using data recorded with the Belle detector operated at the KEKB asymmetric-energy 
$\EE$ collider~\cite{KEKB}. Charge-conjugated modes are implicitly included throughout the paper. For the $\ytwos$ 
data sample, we have collected data corresponding to an integrated luminosity of $24.7~\infb$ at a c.m. energy 
corresponding to the $\ytwos$ resonance. We determine the number of produced $\ytwos$ events to be $(158\pm 4)\times 
10^6$ using inclusive hadronic decays. The continuum production of the various final states is based on an 
off-resonance data sample collected using an integrated luminosity of $89.5~\infb$ at $\sqrt{s} = 10.52~\gev$. We use 
these two data samples to separate the dynamics of electromagnetic and strong charmed 
hadron production at the off-resonance energy and the $\ytwos$ peak.

We only include the following $\dsj$ states, which are both established and emit a kaon in their decay: $\dsa$ and 
$\dsb$; the kaon can be either charged or neutral ($\ks$). We use the technique of partial reconstruction for the 
$\dsj$ final state: the final state is tagged through the full reconstruction of the $\dsstp$, and the recoiling 
$\dsj$ is tagged by a kaon produced in the decay $\dsj \to \kbar + \dmbar$. The remaining $\dmbar$ is observed 
indirectly through its recoil against the $\dsstp-\kbar$ system using the known kinematics of the initial state. This 
circumvents the problem of low efficiencies for the reconstruction of $D$ mesons associated with the large variety of 
possible decay processes. 

\section{The Belle detector and Monte Carlo simulation}

The Belle detector is a large-solid-angle magnetic spectrometer~\cite{Belle} using a silicon vertex detector, a 
50-layer central drift chamber, an array of aerogel threshold Cherenkov counters, a barrel-like arrangement of 
time-of-flight scintillation counters, and an electromagnetic calorimeter (ECL) comprised of CsI(Tl) crystals located 
inside a super-conducting solenoid coil that provides a 1.5 T magnetic field. An iron flux return located outside of 
the coil is instrumented to detect $\kl$ mesons and to identify muons. The origin of the coordinate system is defined 
as the position of the nominal interaction point. In the cylindrical coordinates, the $z$ axis is aligned with the 
direction opposite the $e^{+}$ beam and points along the magnetic field within the solenoid, and $r$ is the radial 
distance.

We simulate the full chain $\ytwos/\EE \to \dsstp \dsj$, in which $\dsj$ is $\dsa$ or $\dsb$, using the EvtGen 
generator~\cite{Gen}. We simulate the angular distributions of $\dsstp\dsj$ according to the $J^P$ quantum numbers of 
$\dsstp$ and $\dsj$. Here, we take $J^P = 1^-$ for $\dstp$ according to the recent BESIII measurement~\cite{bes3}. 
Four decay modes of $\dsj$ are simulated: $\KM + \adz$, $\ks + \dm$, $\KM + \adza$, and $\ks + \dzb$. Again, 
the $D$ mesons ($\adz$, $\dm$, $\adza$, and $\dzb$) are not reconstructed, but determined in the recoil of 
the $\dsstp$ and the kaon from the $\dsj$ decay, so that the decays of $D$ mesons are inclusive. We simulate the 
response of the Belle detector using a GEANT3-based Monte Carlo (MC) technique~\cite{GSIM}. 

\section{Event Selection Criteria and Reconstruction}

We search for the tagging $\dsp$ using six final states: $\phi\pip$, $\ks\kap$, $\ktb\kap$, $\rhop\phi$, $\eta\pip$ 
and $\etap\pip$. The decay of $\dstp$ only proceeds through $\dstp \to \dsp \gamma$.

We reconstruct $\dsstp\dsj$ final states by first selecting well-measured charged tracks and photon candidates. A 
well-measured charged track has an impact parameter $dr < 1.5~\cm$ in the $r-\phi$ plane with 
respect to the interaction point and a displacement $|dz|<5~\cm$ in the $r-z$ plane. We require a transverse momentum 
larger than $0.1~\gevc$. We identify each charged track by combining the information from different detector 
subsystems and form the likelihood $\mathcal{L}_i$ \cite{pid} for each particle species $i$, denoting $\pi$ or $K$. 
Tracks with $\mathcal{R}_K = \frac{\mathcal{L}_K} {\mathcal{L}_K+\mathcal{L}_\pi} > 0.6$ are treated as kaons, while 
those with $\mathcal{R}_K < 0.4$ are assumed to be pions. The identification efficiency is about 95\% for both $K$ 
and $\pi$. Photons are identified through a cluster in the ECL, which does not align with any charged track.

We reconstruct the $\ks$, $\phi$, $\ktb$, and $\rhop$ candidates in their respective decay channels into 
$\pp$, $\kk$, $\kam\pip$, and $\pip\piz$. For candidate $\ks$ mesons, we use pairs of oppositely charged 
particles that originate from a common vertex and assign the pion-mass hypothesis. We use a multivariate technique to 
improve the purity of the $\ks$ candidate sample by rejecting combinatorial background~\cite{ks01}, which we identify 
with neural network (NN)~\cite{ks02} based algorithms. For the invariant mass ($M_{\pp}$) of $\pp$ pairs we obtain a 
resolution of $\sigma \approx 5~\mevcs$, and we define the signal region for $\ks$ by $ |M_{\pp} - m_{\ks}|< 
3\sigma$. Here, $m_{\ks}$ is the nominal mass of the $\ks$~\cite{PDG}. Correspondingly, we choose the 
range of all 
signal mass windows to have $\Delta m = \pm 3\sigma$ around their respective nominal masses~\cite{PDG}, unless stated 
otherwise. The corresponding resolution for the invariant mass of $\kk$ pairs, $M_{\kk}$, is $\sigma \approx 
3.3~\mevcs$. The $\ktb$ meson has a natural width of $47.3~\mev$, which is much larger than the experimental 
resolution for $\kam\pip$ pairs. We define the $\ktb$ signal region to be $|M_{\kam\pip} - m_{\ktb}| < 105~\mevcs$, 
where $M_{\kam\pip}$ is the invariant mass of $\kam\pip$ and $m_{\ktb}$ is the nominal mass of $\ktb$~\cite{PDG}. 
Since the width of the $\rhop$ of about $150~\mev$ is dominated by the natural width, the signal 
region is selected by $|M_{\pip\piz}-m_{\rhop}| < 200~\mevcs$, in which $M_{\pip\piz}$ is the invariant mass of $\pip 
\piz$ and $m_{\rhop}$ is the nominal mass of $\rhop$~\cite{PDG}. 

We combine pairs of photons to form  $\piz$ candidates. For this, we require the energies of photons ($E_\gamma$) 
from $\piz$ decays to exceed $E_\gamma > 25~\mev$ in the barrel ($32.2^\circ < \theta < 128.7^\circ$) and $E_\gamma > 
50~\mev$ in the endcaps ($12.0^\circ < \theta < 31.4^\circ$ or $131.5^\circ < \theta < 157.1^\circ$) of the ECL, with 
the polar angles specified in the laboratory frame. The two-photon mass resolution for $M_{\GG}$ is $\sigma \approx 
5~\mevcs$. We reconstruct $\eta$ mesons from their decay into both $\ppp$ and $\GG$. In the $\GG$ mode, we require 
$E_\gamma > 150~\mev$. The corresponding mass resolution for $M_{\ppp}$ is $\sigma \approx 4~\mevcs$ and for 
$M_{\GG}$ the resolution is $\sigma \approx 13.5~\mevcs$. For the selection of $\etap$ candidates, we use a 
combination of $\eta$ and $\pp$ pairs. The invariant mass resolution for $\eta\pp$ is $\sigma_{\eta\pp}\approx 
5~\mevcs$. 

\begin{figure}[tbp]
 \psfig{file=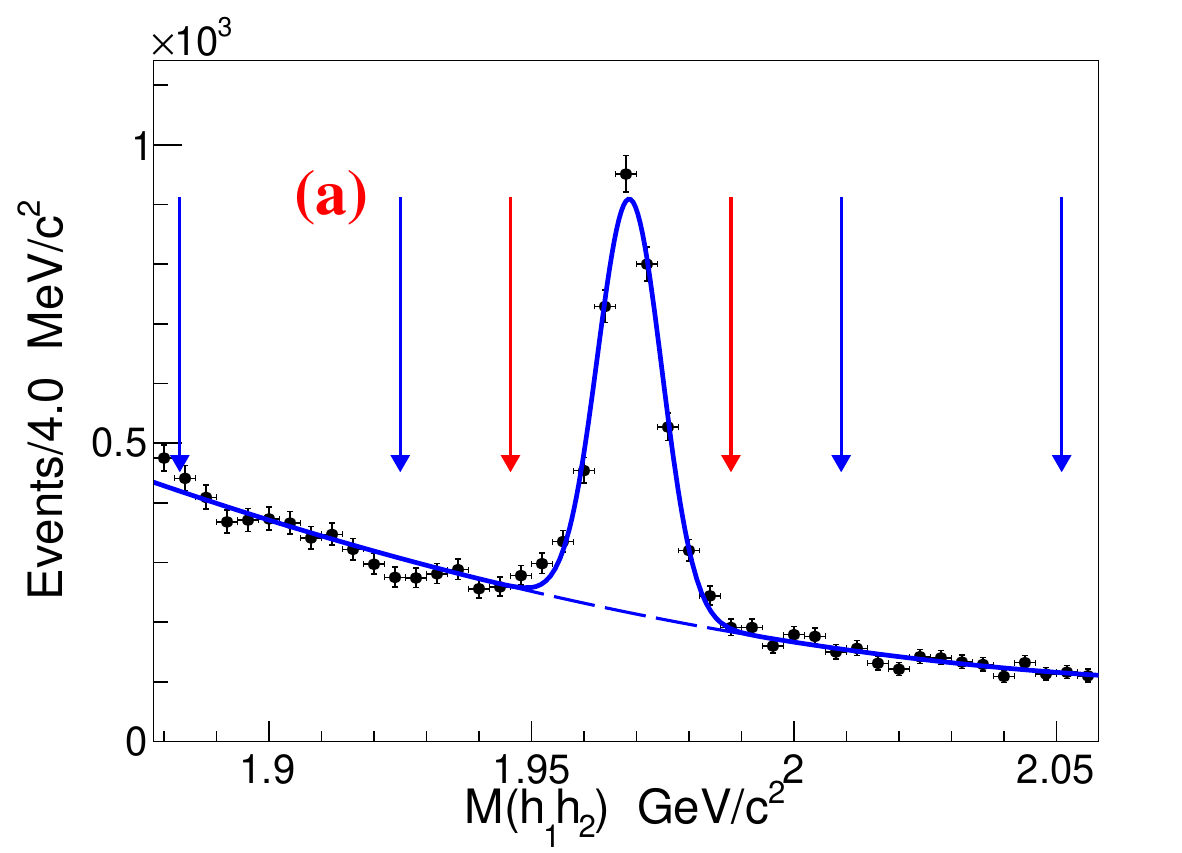, width=0.45\textwidth}
 \psfig{file=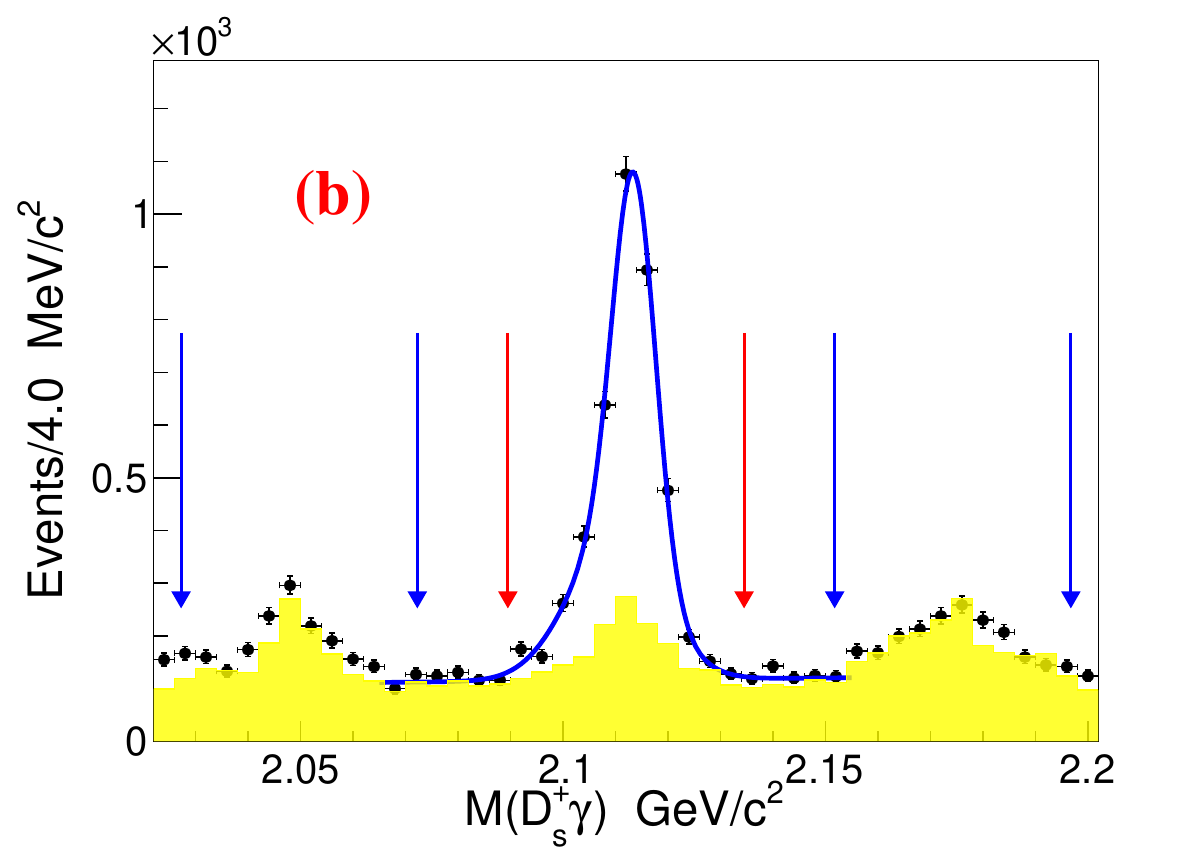, width=0.45\textwidth}\\
 \psfig{file=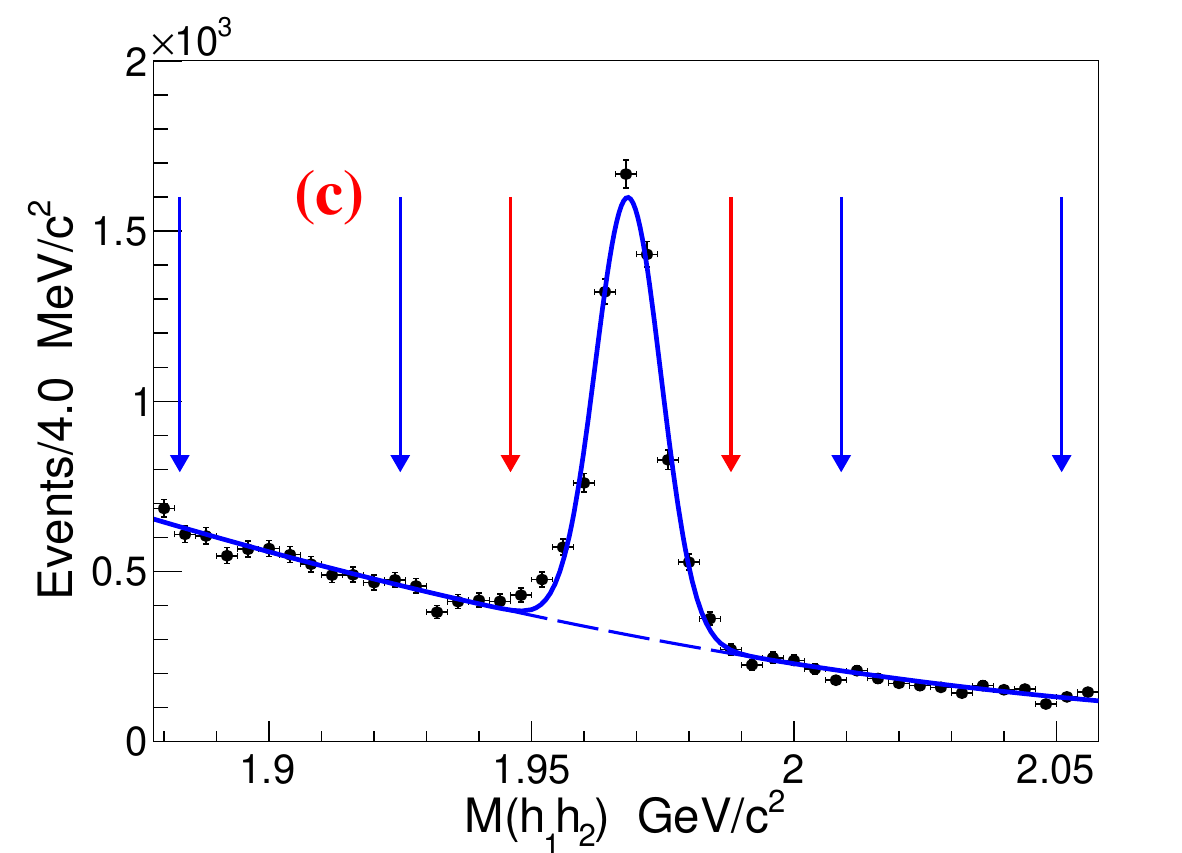, width=0.45\textwidth}
 \psfig{file=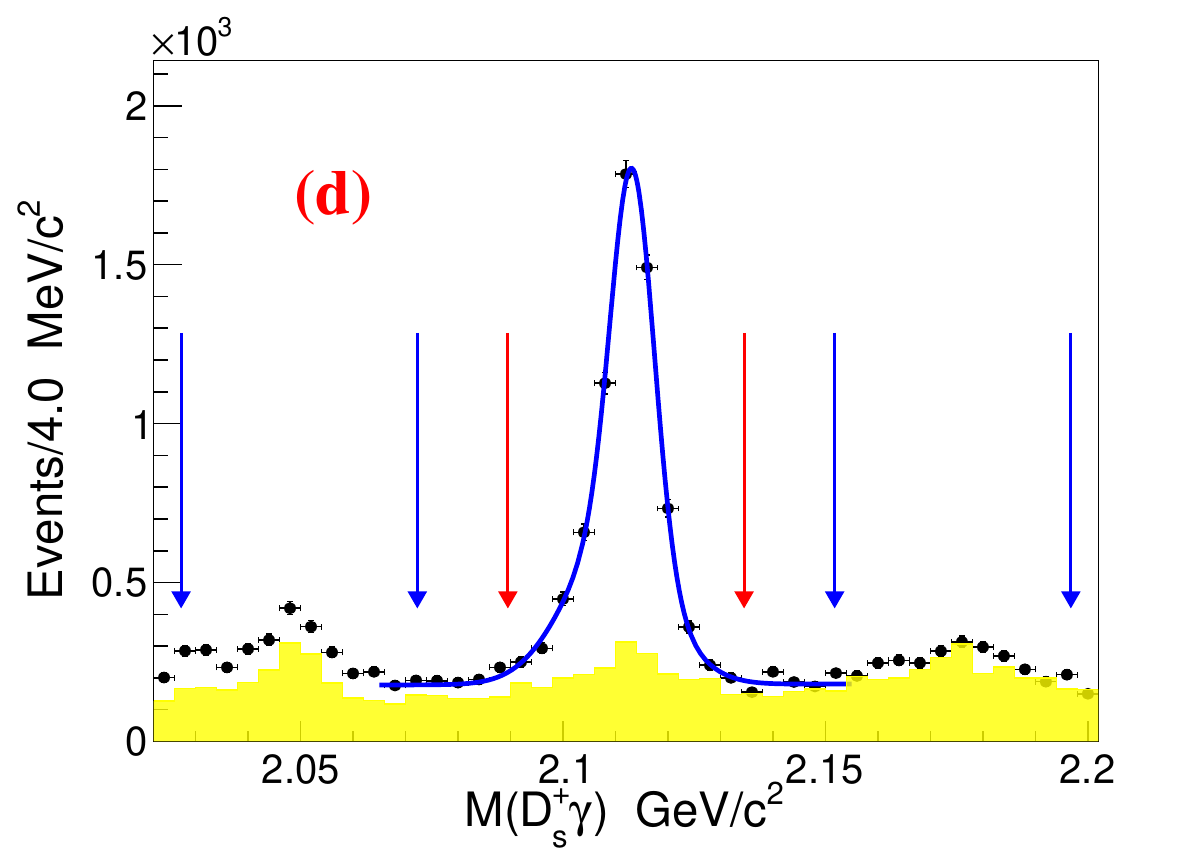, width=0.45\textwidth}
\caption{Invariant mass distributions of (a,c) the combinations of $\phi\pip$, $\ks\kap$, $\ktb\kap$, $\rhop\phi$,
$\eta\pip$, and $\etap\pip$ for $\dsp$ candidates and (b,d) the combinations of $\gamma\dsp$ for $\dsstp$ 
candidates in $\ytwos$ data sample (upper panels) and continuum data sample (lower panels). The red arrows show the 
signal region of $\dsp$ or $\dsstp$ and the blue arrows show the related sideband regions. The shaded histogram in 
(b) and (d) shows backgrounds estimated from $\dsp$ mass sidebands. The curves show the best fit results using 
Gaussian functions for the $\dsp$ and $\dsstp$ signals, respectively.} 
\label{DSST}
\end{figure}

In Fig.~\ref{DSST}(a) and (c), we show the combined distribution $M_{\rm h_1 h_2}$ of $M_{\phi\pip}$, $M_{\ks\kap}$, 
$M_{\ktb\kap}$, $M_{\rhop\phi}$, $M_{\eta\pip}$ and $M_{\etap\pip}$ from the $\ytwos$ data sample (upper row) and the 
continuum data sample (lower row). We do not apply a mass constraint for $\piz$, $\eta$ or 
$\eta^\prime$. Instead, 
we take the advantage of the mass difference. Taking the $\dsp \to \eta \pip$ with $ \eta \to \GG$ as an example, we 
use $M_{\eta\pip} = M_{\GG\pip} - M_{\GG} + m_\eta$, where the invariant mass $M_{\GG\pip}$ ($M_{\GG}$) is calculated 
from the sum of the 4-momenta of $\GG\pip$ ($\GG$). In this way, the mass resolution of the $\dsp$ signal in 
$M_{\eta\pip}$ is improved from $19.7~\mevcs$ to $13.0~\mevcs$ according to signal MC simulation. We fit the 
$\dsp$ signal in $M_{\rm h_1 h_2}$ with a Gaussian function and describe the 
background through a second-order polynomial function. We obtain a mass resolution of $\sigma_{\dsp} = 6.7 
\pm 0.1~\mevcs$ in data, which is used to define the signal region for $\dsp$, while the corresponding resolution is 
$6.5~\mevcs$ in signal MC simulations. Besides the $\dsp$ signal, we also define sideband regions through 
$|M_{h_1 h_2} - m_{\dsp} \pm 9\sigma_{\dsp}|< 3\sigma_{\dsp}$. Since the fraction of multi-combination in $\dsp$ 
reconstruction is only about 3\%, we allow multiple candidates of $\dsp$ in one event. 

We reconstruct $\dstp$ candidates from the above $\dsp$ sample using the $\gamma\dsp$ final state. For this, we 
require the photon energy to exceed $E_\gamma > 50~\mev$ in the barrel and $E_\gamma > 100~\mev$ in the endcaps of 
the ECL. The corresponding invariant mass distributions $M_{\gamma\dsp}$ for $\gamma\dsp$ from the $\ytwos$ and 
continuum data samples are shown in Fig.~\ref{DSST}(b) and (d). Here, we use  
$M_{\gamma\dsp} = M_{\gamma h_1 h_2} - M_{h_1 h_2} + m_{\dsp}$, where the invariant mass $M_{\gamma h_1 h_2}$ is 
calculated from the sum of the 4-momenta of $\gamma h_1 h_2$. We fit to the $M_{\gamma\dsp}$ distribution between 
$2.07~\gevcs$ and $2.15~\gevcs$ using two Gaussian functions for the $\dstp$ signal and a second order polynomial 
function for the background. We use $\sigma \equiv \sqrt{f_1\times (\sigma_1^2+m_1^2) + f_2\times (\sigma_2^2+m_2^2) 
- m^2}$ with $m=f_1\times m_1+f_2\times m_2$ to define the mass resolution of the $\dstp$ signals, where $m_1$ 
($m_2$), $\sigma_1$ ($\sigma_2$) and $f_1$ ($f_2$) are the mean, the standard deviation and the fraction of the first 
(second) Gaussian function. We obtain the mass resolution of $\sigma_{\dstp} = 6.7\pm 0.4~\mevcs$ in data and 
$7.0~\mevcs$ in signal MC simulations, which agree well with each other. Again, in addition to the signal region for 
$\dstp$, we define sideband regions through $|M_{\gamma\dsp} - m_{\dstp} \pm 9\sigma_{\dstp}| < 3\sigma_{\dstp}$. As 
we aim to study the $\dstp \kbar$ recoil spectrum, we apply mass-constrained fits to the $\dstp$ candidates in the 
signal region to improve their momentum resolution. We find that 35\% of the events have multiple $\dstp$ candidates. 
In these cases, we select the candidate with the minimum $\chi^{2}$ from the mass-constraint fit. For the candidates 
in each $\dstp$ mass sideband, we apply the mass constraint to the center of the sideband and select the combination 
with minimum $\chi^2$ as well. To estimate the size of the peaking component in the selected $\dstp$ sample due to 
the minimum $\chi^2$ requirement, we apply the same mass constraints to events in the $\dsp$ sidebands. As shown in 
Fig.~\ref{DSST}(b) and (d), the $\dsp$ mass sideband events can describe the peaks in the $\dstp$ mass sidebands, and 
therefore can be used to estimate the peaking component in the $\dstp$ mass signal region reliably. Events with 
$|M_{\gamma\dsp} - m_{\dstp}| < 50~\mevcs$ are removed for the $\dsp \dsj$ search.

\begin{figure}[tbp]
\psfig{file=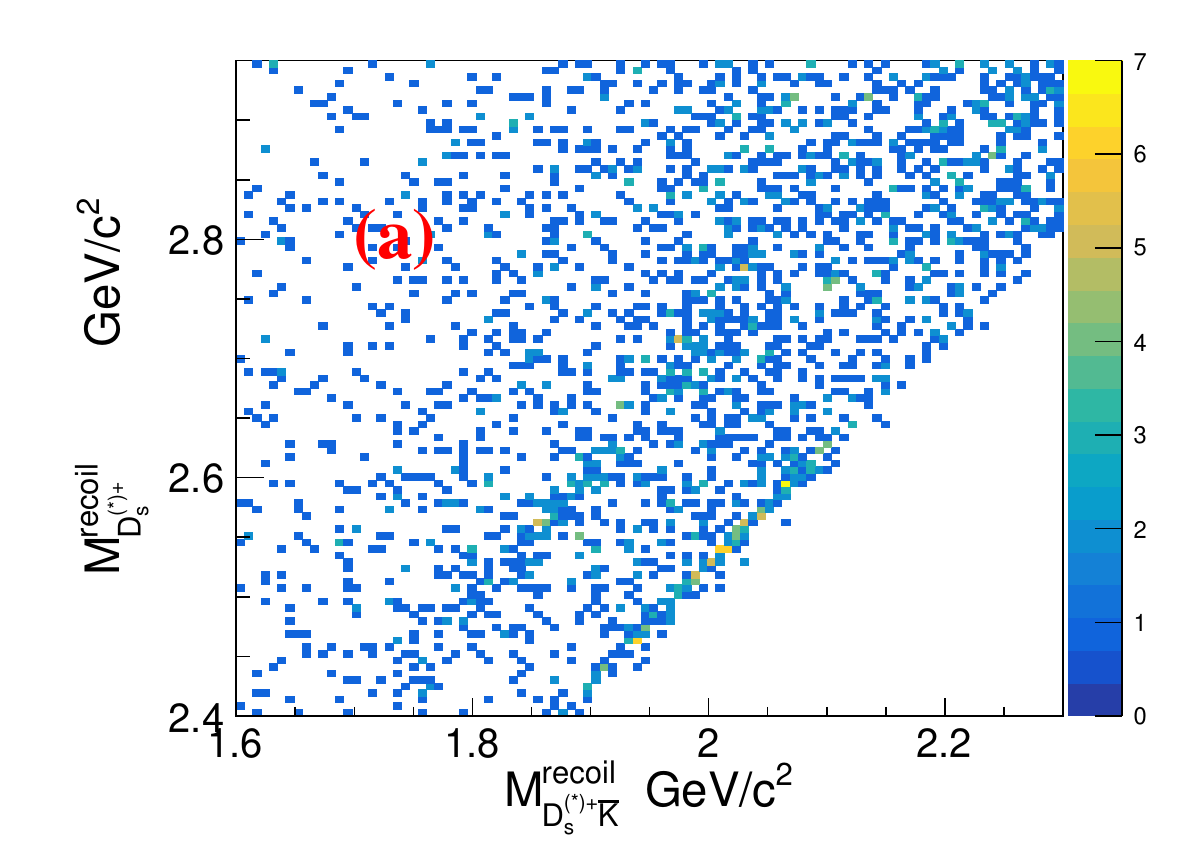, width=0.49\textwidth}
\psfig{file=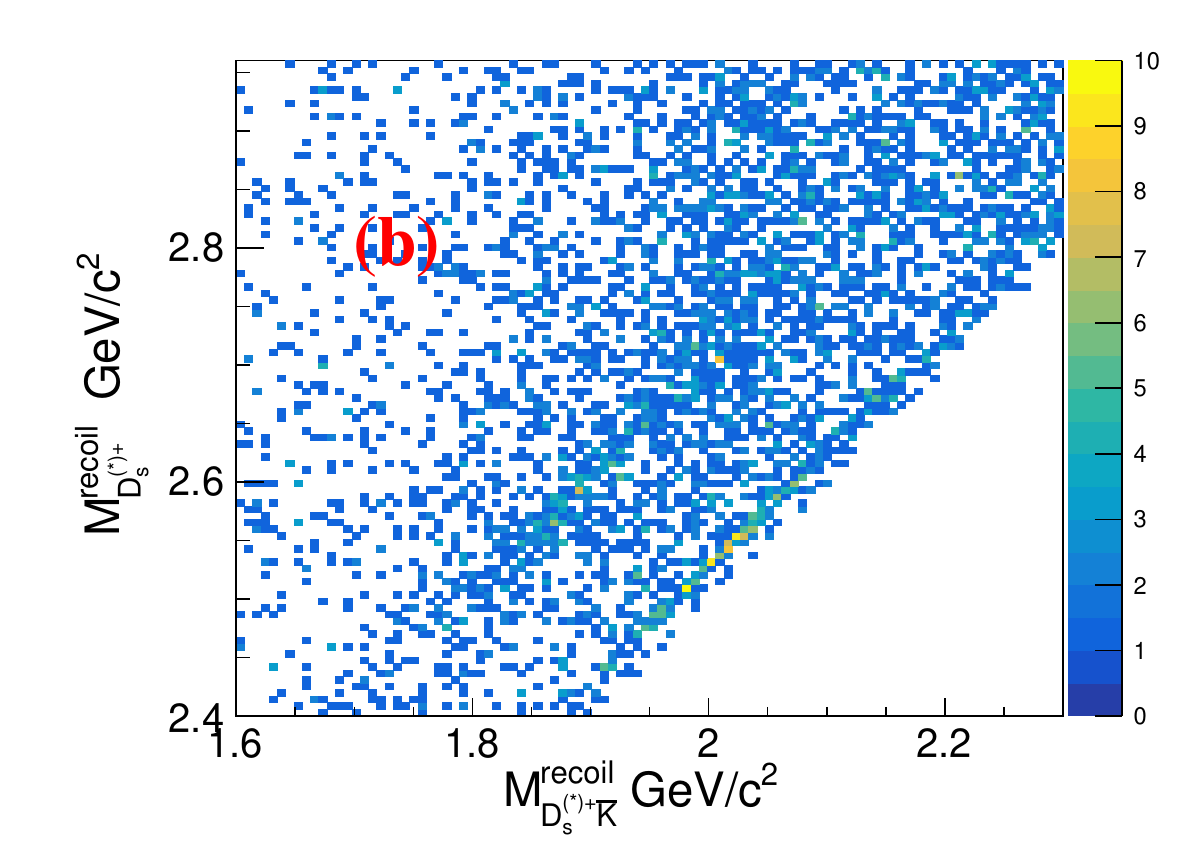, width=0.49\textwidth}\\
\psfig{file=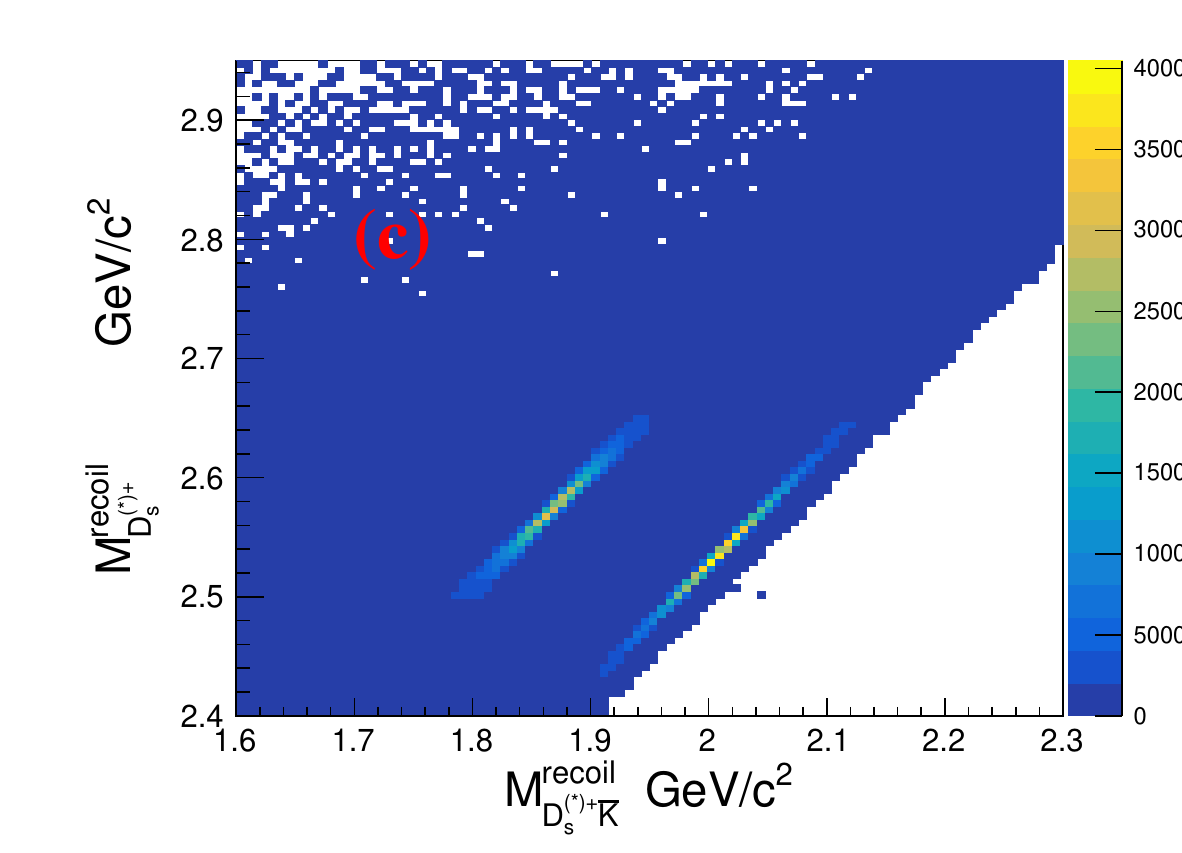, width=0.49\textwidth}
\psfig{file=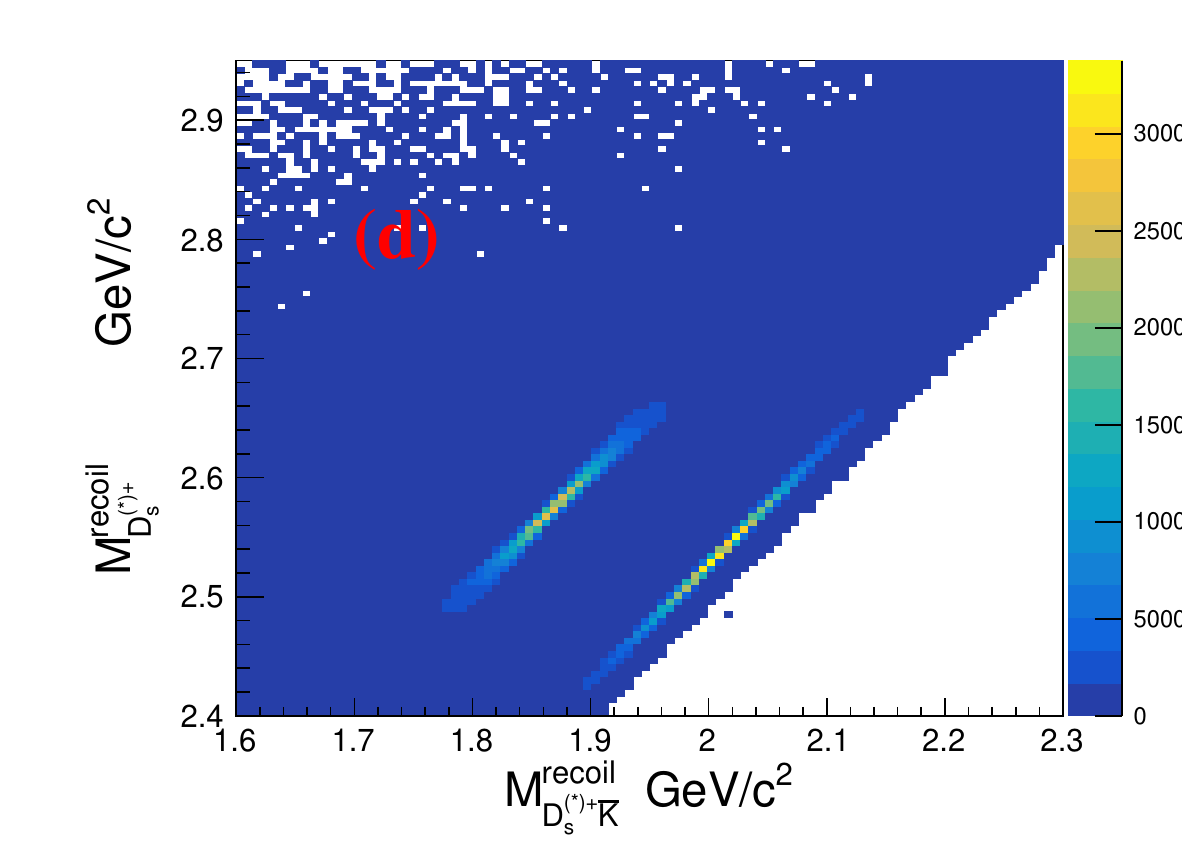, width=0.49\textwidth}
\caption{The distributions of the recoil mass against $\dsstp\kbar$ versus the recoil mass against $\dsstp$ in (a) 
the $\ytwos$ data sample, (b) the continuum data sample at $\sqrt{s} = 10.52~\gev$, (c) the signal MC simulation of 
$\ytwos$ decays and (d) the signal MC simulation of continuum production at $\sqrt{s} = 10.52~\gev$. }
\label{RECDSK}
\end{figure}

The search for $\dmbar$ requires a $\kbar$ meson reconstructed in addition to $\dsstp$. We determine the $\dmbar$ 
signal through the recoil of $\dsstp\kbar$ using the calculated mass:
\beq\label{eq_2}
M_{\dmbar} = \mrec_{\dsstp \kbar}\equiv \sqrt{(\ecm -E_{\dsstp}-E_{\kbar})^2 - (\vec{p}_{\rm c.m.} - 
\vec{p}_{\dsstp}-\vec{p}_{\kbar})^2},
\eeq
and isolate the possible production of $\dsj$ states in the $\kbar \dmbar$ final states through their recoil using 
the following equation: 
\beq\label{eq_1}
M_{\kbar \dmbar} = \mrec_{\dsstp} \equiv \sqrt{(\ecm - E_{\dsstp})^2 - (\vec{p}_{\rm c.m.} - \vec{p}_{\dsstp})^2}.
\eeq
Here, $\ecm$ and $\vec{p}_{\rm c.m.}$ are the energy and 3-momentum of $\EE$ in the collision system, $E_{\dsstp}$ 
($E_{\kbar}$) and $\vec{p}_{\dsstp}$ ($\vec{p}_{\kbar}$) are those of $\dsstp$ ($\kbar$), respectively. We show the 
$\mrec_{\dsstp\kbar}$ distributions versus $\mrec_{\dsstp}$ from the two data samples in Fig.~\ref{RECDSK}(a) and 
(b), and the signal MC simulations of $\ytwos$ decays and continuum productions in Fig.~\ref{RECDSK}(c) and (d). 
There are clear bands in the distributions of data corresponding to the production of the $\dsa$ signal in the 
$\dtbar\kbar$ ($\adza\kam$ or $\dzb\ks$) final state and $\dsb$ signal in the $\dbar\kbar$ ($\adz\kam$ or $\dm\ks$) 
final state, and they agree well with the signal MC simulations. The mass resolutions of $\mrec_{\dsstp \kbar}$ and 
$\mrec_{\dsstp}$ are large due to the common variables $E_{\dsstp}$ and $\vec{p}_{\dsstp}$ in Eqs.~(\ref{eq_2}) and 
(\ref{eq_1}). The mass resolution of $\dbar$ from the decay of $\dsj$ in 
$M^{\rm recoil}_{\dsstp\kbar}$ is about $50~\mevcs$, and the signal region is defined to be $|\mrec_{\dsstp \kbar} - 
m_{\dbar}|<150~\mevcs$. We fit to the $\dtbar$ mass distribution with two Gaussian functions, and obtain the narrower 
one with a mass resolution of $31.8\pm 0.3~\mevcs$ and a signal fraction of about 34\% and the wider one with a mass 
resolution of $74.2\pm 1.0~\mevcs$ and a signal fraction of about 66\%. We define the signal region to be 
$|\mrec_{\dsstp \kbar} - m_{\dtbar}| < 200~\mevcs$, which has a selection efficiency of about 95\%. Here, $m_{\dbar}$ 
is the nominal mass of $\adz$ or $\dm$, and $m_{\dtbar}$ is the nominal mass of the $\adza$ or $\dzb$~\cite{PDG}. 
With the events in the $\dsp$ or $\dstp$ mass sidebands, no peaking background is found for the $\dmbar$ signal in 
the $\mrec_{\dsstp\kbar}$ distributions.

\begin{figure}[tbp]
\psfig{file=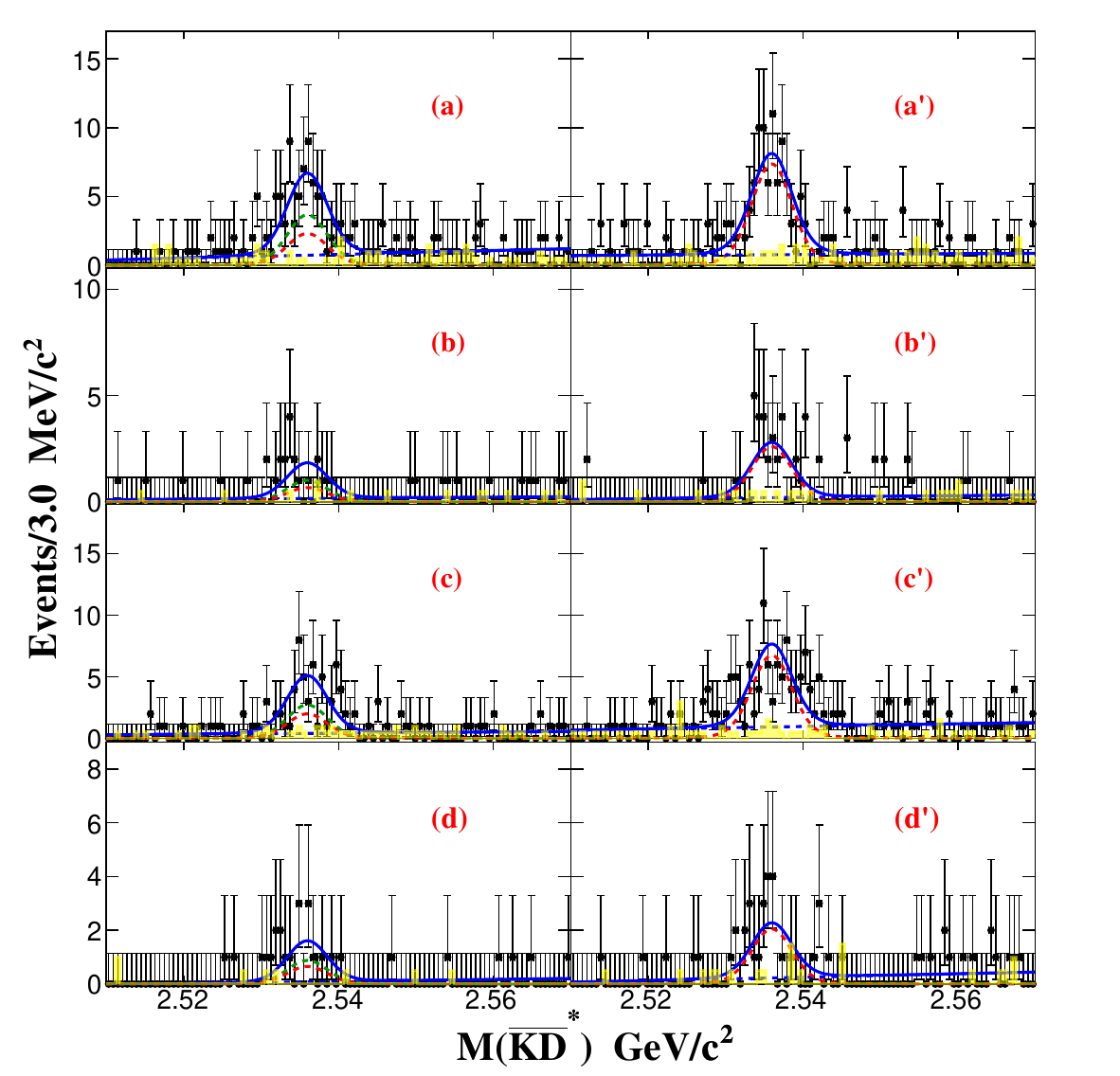, width=0.8\textwidth}
\caption{The invariant mass distributions of $\kbar\dtbar$ calculated in the recoil mass for $\dsstp$ in the (a) 
$\dsp\KM\adza$, (b) $\dsp\ks\dzb$, (c) $\dstp\KM\adza$, and (d) $\dstp\ks\dzb$ final states from the $\ytwos$ 
data sample (left panels) and the continuum data sample at $10.52~\gev$ (right panels). The shaded histograms show 
the backgrounds estimated from the normalized $\dsstp$ mass sidebands. The solid curves show the best fit result; the 
dashed green ones are $\dsa$ signals in $\ytwos$ decays and the dashed red curves are the $\dsa$ signals in continuum 
production at $10.02~\gev$ (left panels) and $10.52~\gev$ (right panels). }
\label{dsa_sigs}
\end{figure}

\begin{figure}[tbp]
\psfig{file=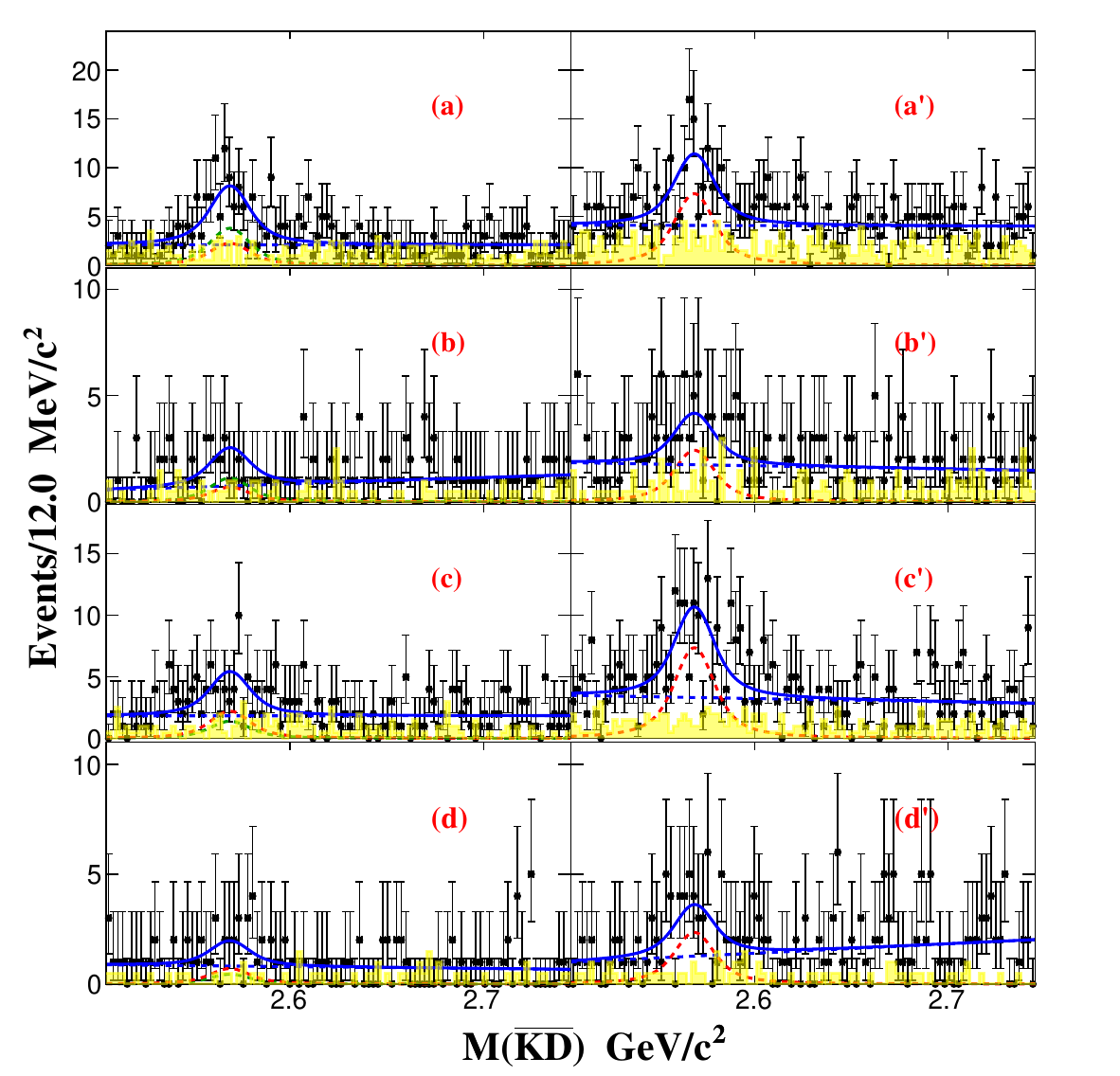, width=0.8\textwidth}
\caption{The invariant mass distributions of $\kbar\dbar$ calculated in recoil mass for $\dsstp$ in the (a) 
$\dsp\KM\adz$, (b) $\dsp\ks\dm$, (c) $\dstp\KM\adz$, and (d) $\dstp\ks\dm$ final states from the $\ytwos$ 
data sample (left panels) and the continuum data sample at $10.52~\gev$ (right panels). The shaded histograms show 
the backgrounds estimated from the normalized $\dsstp$ mass sidebands. The solid curves show the best fit result; the 
dashed green ones are $\dsb$ signals in $\ytwos$ decays and the dashed red curves are the $\dsb$ signals in continuum 
production at $10.02~\gev$ (left panels) and $10.52~\gev$ (right panels). }
\label{dsb_sigs}
\end{figure}

To improve the mass resolution of $M_{\kbar \dmbar}$, we use the following formula to replace Eq.~(\ref{eq_1}):
\beq\label{eq_3}
M_{\kbar \dmbar} = \mrec_{\dsstp} - \mrec_{\dsstp \kbar} + m_{\dmbar}.
\eeq
In this way, the uncertainties due to the 4-momentum of final states from $\dsstp$ decays are significantly reduced. 
From simulation, we obtain the resolutions for $\Delta\mrec\equiv \mrec_{\dsstp} - \mrec_{\dsstp \kbar}$ of 
$\sigma_{\Delta \mrec}<5~\mevcs$ for all $\dsstp\dsj$ final states. In Figs.~\ref{dsa_sigs} and \ref{dsb_sigs} we 
show the distributions for $\Delta \mrec + m_{\dtbar}$ for $M_{\kbar\dtbar}$ and $\Delta \mrec + m_{\dbar}$ for 
$M_{\kbar\dbar}$ for the two data samples. We observe clear signals for both $\dsa$ and $\dsb$.

We determine the numbers of $\dsj$ signals, $N^{\rm sig}_{\ytwos}$ of the $\ytwos$ decays and 
$N^{\rm sig}_{\rm cont}$ of the continuum production at $\sqrt{s} = 10.52~\gev$, by simultaneously fitting the 
$M_{\kbar\dmbar}$ distributions for the $\ytwos$ data sample and the continuum data sample, and with common isospin 
ratios $R_{{\rm iso},J} \equiv \BR(\dsj\to \ks \dmab)/\BR(\dsj\to \kam \adzab)$ between the $\ks\dmab$ and $\kam 
\adzab$ final states. In the fits, we use $N^{\rm sig}_{\ytwos}$ and $N^{\rm sig}_{\rm cont}$ of the $\kam\adzab$ 
modes, and those of the $\ks\dmab$ modes are calculated via the isospin ratios $R_{{\rm iso},J}$ and the ratios of 
efficiencies and branching fractions between the $\ks\dmab$ modes and the $\kam\adzab$ modes. The fit function is the 
sum of a Breit-Wigner function (BW) convoluted with a Gaussian function with a width corresponding to the mass 
resolution, and a linear function to describe the backgrounds. The mass and width of the BW functions are fixed to 
the world average values for $\dsa$ and $\dsb$~\cite{PDG}. The mass resolutions used in the Gaussian are obtained 
from MC simulations and are about $2.4~\mevcs$ ($6.5~\mevcs$) for $\dsa$ ($\dsb$). In the fits, we include the 
branching fractions and reconstruction efficiencies corresponding to the $\dsstp\dsj$ final states. The results are 
listed in Table~\ref{br_xs_sum} for each channel and each data set.

We estimate the contribution of continuum production to the $\dsstp\dsj$ signal in the $\ytwos$ data sample. For 
this, we scale the luminosities and correct for the c.m. energy dependence of the QED cross section $\sigma_{\EE} 
\propto 1/s$, resulting in a scale factor $f_{\rm scale}=(\mathcal{L}_{\ytwos}\times s_{\rm cont})/(\mathcal{L}_{\rm 
cont}\times s_{\ytwos}) = 0.304$. Here, $\mathcal{L}_{\ytwos}$ and $\mathcal{L}_{\rm cont}$ are the integrated 
luminosities of the $\ytwos$ data sample at $\sqrt{s_{\ytwos}} = 10.02~\gev$ and the continuum data sample at 
$\sqrt{s_{\rm cont}} = 10.52 ~\gev$. Therefore, the yield of signal events produced via continuum $\EE$ annihilation 
in the $\ytwos$ data sample is $f_{\rm scale} \times N^{\rm sig}_{\rm cont}$.

We determine the statistical significance of $\dsj$ by comparing the value of $\Delta (-2\ln L) = 
-2\ln(L_{\rm max}/L_0)$ and the change in the number of free parameters in the fits, where $L_{\rm max}$ is the 
likelihood with $\dsj$ and $L_0$ without $\dsj$. The statistical significance in the $\ytwos$ data sample for $\dsa$ 
and $\dsb$ is $6.8\sigma$ and $4.0\sigma$, respectively, and $18.3\sigma$ and $10.1\sigma$ in the continuum 
data sample. From these yields, we calculate the branching fraction of $\ytwos \to \dsstp \dsj$ and the Born cross 
section for $\EE\to \dsstp \dsjm$ by
\beq
\label{eq_br}
\BR(\ytwos \to \dsstp \dsj)\BR(\dsj\to \kbar \bar{D}^{(*)}) 
= \frac{N^{\rm sig}_{\ytwos}}{N_{\ytwos}\times \sum \eff_i \BR_i},
\eeq
and
\beq\label{eq_xs}
\sigma^{\rm B}(\EE\to \dsstp\dsj)\BR(\dsj\to \kbar \bar{D}^{(*)}) = \frac{N^{\rm sig}_{\rm cont}\times |1-\Pi|^2}
{\mathcal{L}_{\rm cont}\times \sum \eff_i \BR_i \times (1+\delta_{\rm ISR})}.
\eeq
Here, $i$ identifies the mode of $\dsp\to h_1h_2$ decay, while $\eff_i$ and $\BR_i$ are their reconstruction 
efficiencies and branching fractions. We calculate $\sum \eff_i \BR_i$ according to signal MC simulations for 
$\eff_i$ and the world average values of $\BR_i$~\cite{PDG}. In the $\dsstp\dsa$ channels, they are $(1.63\pm 
0.07)\%$, $(1.06\pm 0.05)\%$, $(1.19\pm 0.05)\%$, and $(0.77\pm 0.03)\%$ in the final states of $\dsp\kam \adza$, 
$\dsp\ks\dzb$, $\dstp\kam\adza$, and $\dstp\ks\dzb$, respectively. In the $\dsstp\dsb$ channels, they are $(2.32\pm 
0.10)\%$, $(1.56\pm 0.07)\%$, $(1.22\pm 0.05)\%$, and $(0.82\pm 0.03)\%$ in the final states of $\dsp\kam \adz$, 
$\dsp\ks\dm$, $\dstp\kam\adz$, and $\dstp\ks\dm$, respectively. The errors of these values are mainly due to the 
uncertainties of $\BR_i$ from world averages~\cite{PDG}, since all of the relative statistical uncertainties due to 
the statistics of MC simulations are less than 0.5\%. Here, we take into account the branching fraction of $\ks\to 
\pp$ decay~\cite{PDG}. From the Born cross-sections we can calculate the full ``dressed'' cross section through 
$\sigma^{\rm dressed} = \sigma^{\rm Born}/|1-\Pi|^2$. The factor $|1-\Pi|^2 = 0.931$ is the vacuum polarization 
factor~\cite{vacuum, vacuum_2}. In addition, we have to correct for radiative effects. The radiative correction 
factor $1+\delta_{\rm ISR}$ is determined by $\int \sigma^{\rm dressed}(s(1-x)) F(x,s)dx/\sigma^{\rm dressed}(s)$ and 
has the value 0.82, where $F(x,s)$ is the radiative function obtained from a QED calculation with an accuracy of 
0.2\%~\cite{radia_correction, radia_correction_2, radia_correction_3}. 

We summarize the branching fractions of $\ytwos$ decays and the Born cross sections of continuum production in 
Table~\ref{br_xs_sum}. The number of corrected signal events in the $\ytwos$ data sample is $20\pm 12\pm 2$ for the 
$\dstp\dsb$ decay, from which we derive a statistical significance of only $1.6\sigma$. We integrate the likelihood 
versus the number of $\dstp\dsb$ signal events, and determine its upper limit at 90\% confidence level (C.L.) to be 
$N^{\rm UL}(\ytwos\to \dstp\dsb) < 44$ in the $\dsb\to \kam\bar{D}^0$ mode, which has been degraded by a factor of 
$1/(1-\delta_{\rm sys})$ to account for the systematic uncertainties detailed below. We obtain an upper limit for the 
production in $\ytwos$ decay of $\BR^{\rm UL}(\ytwos\to \dstp\dsb)\BR(\dsb\to \kam\bar{D}^0) < 2.5\times 10^{-5}$. 

\begin{table}
\caption{The branching fractions of $\ytwos\to\dsstp\dsj$ decays and the Born cross sections of continuum production 
$\EE\to\dsstp\dsj$ based on the results from the simultaneous fits. Here, $N^{\rm sig}_{\ytwos}$, $N^{\rm sig}_{\rm 
cont}$, $\BR(\ytwos \to \dsstp \dsj) \BR(\dsj\to \kbar \bar{D}^{(*)})$, and $\sigma^{\rm B}(\EE\to \dsstp
\dsj)\BR(\dsj\to\kbar\bar{D}^{(*)})$ are described in Eqs.~(\ref{eq_br}) and (\ref{eq_xs}). The significance is the 
statistical significance of the $\dsstp\dsj$ signals with $\dsj\to \kam \dzab$ and $\ks \dmab$ in $\ytwos$ decays and 
continuum productions. The $\kam \dzab$ and $\ks \dmab$ modes of the $\dsj$ decays are connected by the isospin ratio 
$\BR(\dsj\to \ks \dmab)/\BR(\dsj\to \kam\adzab)$ in the simultaneous fits. The systematic uncertainties of 
$N^{\rm sig}$ are of the simultaneous fits only.}
\label{br_xs_sum}
\begin{center}
\renewcommand{\arraystretch}{1.2}
\begin{tabular}{c | c  c | c | c  c  }
\hline
\multirow{2}{*}{Final state ($f$)}  & \multicolumn{2}{c|}{$N^{\rm sig}_{\ytwos}$} & Significance 
 & \multicolumn{2}{c}{$\BR(\ytwos \to \dsstp \dsj)\BR(\dsj\to \kbar \bar{D}^{(*)})$}  
\\\cline{2-3}\cline{5-6}
    & $\kam$ mode & $\ks$ mode & ($\sigma$)  & $\kam$ mode & $\ks$ mode \\\hline
$\dsp\dsa$  & $43\pm 9 \pm 2$ & $14 \pm 3 \pm 2$ & 5.3 & $1.6 \pm 0.3\pm 0.2$ & $0.84\pm 0.18\pm 0.14$  \\
$\dstp\dsa$ & $31\pm 8 \pm 2$ & $10 \pm 3 \pm 2$ & 4.3 & $1.4 \pm 0.4\pm 0.2$ & $0.82\pm 0.25\pm 0.18$ \\
$\dsp\dsb$  & $51\pm 15\pm 5$ & $17 \pm 5 \pm 5$ & 3.8 & $1.4 \pm 0.4\pm 0.2$ & $0.69\pm 0.20\pm 0.21$  \\
$\dstp\dsb$ & $20\pm 12\pm 2$ & $7 \pm 4 \pm 4$  & 1.6 & $0.9 \pm 0.5\pm 0.1$ & $0.54\pm 0.31\pm 0.39$ \\\hline
\multirow{2}{*}{---}  & \multicolumn{2}{c|}{$N^{\rm sig}_{\rm cont}$} & \multirow{2}{*}{---} & \multicolumn{2}{c}{$\sigma^{\rm B}(\EE\to \dsstp\dsj)\BR(\dsj\to \kbar \bar{D}^{(*)})$} \\\cline{2-3}\cline{5-6}
    & $\kam$ mode & $\ks$ mode &   & $\kam$ mode & $\ks$ mode \\\hline
$\dsp\dsa$  & $86\pm 10\pm 2$ & $28 \pm 4 \pm 2$ & 13.9 & $67 \pm 8 \pm 6$   & $34\pm 5\pm 4$ \\
$\dstp\dsa$ & $79\pm 10\pm 2$ & $25 \pm 4 \pm 2$ & 11.8 & $84 \pm 11 \pm 11$ & $41\pm 6\pm 6$ \\
$\dsp\dsb$  & $102\pm 17\pm 21$ & $33 \pm 8\pm 5$ & 7.1 & $56\pm 9 \pm 13$   & $27\pm 6\pm 5$ \\
$\dstp\dsb$ & $102\pm 16\pm 6$ & $33 \pm 7 \pm 4$ & 7.6 & $106\pm 17 \pm 12$ & $51\pm 11\pm 9$ \\\hline
\multicolumn{6}{c}{Isospin ratio $\BR(\dsj\to \ks \dmab)/\BR(\dsj\to \kam \adzab)$} \\\hline
\multicolumn{3}{c|}{$\dsa$ decays} & \multicolumn{3}{c}{$0.48\pm 0.07\pm 0.02$}\\
\multicolumn{3}{c|}{$\dsb$ decays} & \multicolumn{3}{c}{$0.49\pm 0.10\pm 0.02$}\\
\hline
\end{tabular}
\end{center}
\end{table}

Additionally, we determine the isospin ratios $R_{{\rm iso},J}$ from the simultaneous fits to be 
$R_{{\rm iso},1} = 0.48\pm 0.07 \pm 0.02$ and $R_{{\rm iso},2} = 0.49 \pm 0.10 \pm 0.02$ for the 
$\dsa$ and $\dsb$, respectively. These ratios are in good agreement with the expectation from isospin symmetry, which 
are 0.498 and 0.497 from a calculation taking into account the phase space. Replacing the $N^{\rm sig}_{\ytwos}$ and 
$N^{\rm sig}_{\rm cont}$ of $\kam\adzab$ modes with those of the $\ks\dmab$ modes in the simultaneous fits, and 
calculating those of the $\kam\adzab$ modes with $R_{{\rm iso},J}$ and the ratios of efficiencies and branching 
fractions between the $\kam\adzab$ modes and the $\ks\dmab$ modes, we obtain the new fit results and calculate the 
$\BR(\ytwos \to \dsstp \dsj) \BR(\dsj\to \kbar \bar{D}^{(*)})$ and $\sigma^{\rm B}(\EE\to \dsstp\dsj)\BR(\dsj\to\kbar
\bar{D}^{(*)})$ of the $\dsj \to \ks \dmab$ decay modes, as listed in Table~\ref{br_xs_sum}. We get the same fit 
results of the isospin ratios 
$R_{{\rm iso},J}$ of the $\dsa$ and $\dsb$ decays. We also determine the $N^{\rm UL}(\ytwos\to \dstp\dsb) < 15$ in 
the $\ks\dm$ mode and $\BR^{\rm UL}(\ytwos\to \dstp\dsb)\BR(\dsb\to \ks \dm) < 1.4 \times 10^{-5}$ at 90\% C.L.

\section{Systematic uncertainties}

The determination of the branching fractions in $\ytwos$ decays and the Born cross sections of continuum productions 
are subject to a variety of systematic uncertainties, which are listed in Table~\ref{tab_sys}. The particle 
identification uncertainty for $K^\pm$ is 1.1\% and 0.9\% per $\pi^{\pm}$~\cite{pid}; the uncertainty of the tracking 
efficiency per track is 0.35\% and is added linearly; the photon reconstruction uncertainty is 2\% for each photon. 
The uncertainties of the efficiency of mass window requirements due to data and MC differences in mass resolutions 
for $\piz$, $\ks$, $\ktb$, $\rhop$, $\phi$, $\eta$, and $\etap$ are measured to be 0.2\%, 0.2\%, 1.0\%, 
1.4\%, 0.1\%, 1.7\%, and 0.3\%, respectively. We take the decay branching fractions and their uncertainties 
of the intermediate states $\ktb$, $\eta$, $\rhop$, $\etap$, and of $\dsstp$ from Ref.~\cite{PDG}. We determine the 
efficiency of the $\dsp$ ($\dstp$) mass window to be $(99.9\pm 0.1)\%$ [$(99.8\pm 0.1)\%$] in data and 97.4\% 
(99.5\%) in the simulation, and we attribute a systematic uncertainty of 2.5\% (0.3\%); the differences in these 
numbers for data and simulation reflect the different mass resolutions obtained for both. We determine these total 
uncertainties of $\sum \eff_i \times \BR_i$ to be 3.2\%, 3.2\%, 3.3\%, and 3.4\% in the final states of 
$\dsp\kam\adzab$, $\dsp\ks\dmab$, $\dstp\kam\adzab$, and $\dstp\ks\dmab$, respectively. To estimate the 
systematic uncertainty in the angular distribution of $\dsstp\dsj$, we generate new MC samples uniformly in phase 
space, and half of the efficiency differences are taken to be the systematic uncertainties. We get the systematic 
uncertainties of 6.9\%, 8.5\%, 8.5\%, and 9.2\% for the $\dsp\dsa$, $\dsp\dsb$, $\dstp\dsa$, and 
$\dstp\dsb$, respectively. The uncertainty of the total number of $\ytwos$ events is 2.2\%. The uncertainty in the 
integrated luminosities for the two data samples are 1.4\% and are highly correlated, but they cancel in the scale 
factor. We estimate the systematic uncertainty in determining the $\BR^{\rm UL}(\ytwos\to \dstp\dsb)\BR(\dsb\to \KM 
\bar{D}^0)$ to be $\delta_{\rm sys} = 10.1\%$. Besides those listed in Table~\ref{tab_sys}, there are additional 
systematic uncertainties of the scale factor $f_{\rm scale}$ and the radiative correction factor $1+\delta_{\rm 
ISR}$. By changing $s_{\rm cont}/s_{\ytwos}$ to $[s_{\rm cont}/s_{\ytwos}]^{1.5}$, the value of $f_{\rm scale}$ 
changes from 0.304 to 0.319, and we take 4.9\% to be its systematic uncertainty. By varying the photon energy cutoff 
$50~\mev$ in the simulation of ISR, we determine the change of $1+\delta_{\rm ISR}$ to be 0.01 and take 1.0\% to be 
the conservative systematic uncertainty.

Various systematic uncertainties are considered in the simultaneous fit. We change the fit range from 
$[2.51, 2.57]~\gevcs$ to $[2.51, 2.62]~\gevcs$ for the $\dsa$ signals, and from $[2.50, 2.74]~\gevcs$ to 
$[2.50, 2.79]~\gevcs$ for the $\dsb$ signals. We vary the mass and width of $\dsa$ or $\dsb$ by $1\sigma$ according 
to the world average values~\cite{PDG}. We also change the mass resolutions from the signal MC simulations by 
$1\sigma$, and the systematic uncertainties are found to be negligible. 

\begin{sidewaystable}[tbp]
\begin{center}
\caption{The summary of systematic uncertainties ($\%$) of $\dsstp K$ reconstruction.  }
\begin{small}
\begin{tabular}{c | c | c | c | c | c | c | c | c}\hline
        &\multicolumn{6}{c|}{$\dsp$ reconstruction} & $\kam$ & $\ks$   \\\hline
\diagbox{Source}{$\dsp$ decay mode} & $\phi\pip$  & $\ks\kap$ & $\ktb\kap$ & $\rhop\phi$ & $\eta\pip(\GG/\ppp)$ &
$\etap\pip(\GG/\ppp)$ & $\kam$   & $\ks$   \\\hline
$K$ ID         &  2.20   & 1.10  & 2.20  & 2.20  &  ---        & ---         & 1.10 & ---     \\
$\pi$ ID       &  0.90   & ---   & 0.90  & 0.90  &  0.90/2.70  & 2.70/4.50   & ---  & ---     \\
Tracking       &  1.05   & 1.05  & 1.05  & 1.05  &  0.35/1.05  & 1.05/1.75   & 0.35 & ---     \\
$\ks$ reconstruction
               & ---     & 2.23  & ---   & ---   & ---         & ---         & ---  & 2.23    \\
$\pi^0$ reconstruction
               & ---     & ---   & ---   & 2.25  & 2.25/---    & 2.25/---    & ---  & ---     \\
Photon reconstruction
               & ---     & ---   & ---   & ---   & 4.0/---     & 4.0/---     & ---  & ---     \\
Mass windows of intermediate states
               & 0.07    & 0.20  & 0.97  & 1.44  & 0.23/1.68   & 0.26/1.69   & ---  & 0.20    \\
$\BR$s of intermediate state decays
               &  0.08   & 0.08  & 0.08  & 1.12  & 0.04/0.03   & 0.04/0.03   & ---  & ---     \\
$\dsp$ mass window
               &  0.43   &  0.67 & 0.19  &  0.79 & 1.07        &   1.20      & ---  & ---     \\
$\dstp$ mass window
               &  0.38   &  1.01 & 0.10  & 0.34  & 0.94        & 0.61        & ---  & ---  \\\hline\hline
\diagbox{Source}{Reconstruction mode} & \mc{2}{c|}{$\dsp\kam$} & \mc{2}{c|}{$\dsp\ks$} & \mc{2}{c|}{$\dstp\kam$} &
\mc{2}{c}{$\dstp\ks$} \\\hline
$\dsstp K$ reconstruction     & \mc{2}{c|}{3.2} & \mc{2}{c|}{3.2} & \mc{2}{c|}{3.3} & \mc{2}{c}{3.4}    \\
$\BR(\dstp\to\gamma\dsp)$ & \mc{2}{c|}{---} & \mc{2}{c|}{---} & \mc{2}{c|}{0.7} & \mc{2}{c}{0.7}     \\
Trigger                   & \mc{2}{c|}{1.0} & \mc{2}{c|}{1.0} & \mc{2}{c|}{1.0} & \mc{2}{c}{1.0}      \\
MC statistics             & \mc{2}{c|}{0.2} & \mc{2}{c|}{0.2} & \mc{2}{c|}{0.2} & \mc{2}{c}{0.2}     \\
$N_{\ytwos}$ (luminosity) & \mc{2}{c|}{2.2(1.4)} & \mc{2}{c|}{2.2(1.4)} & \mc{2}{c|}{2.2(1.4)} & \mc{2}{c}{2.2(1.4)}\\
Sum in quadrature         & \mc{2}{c|}{4.0(3.6)} & \mc{2}{c|}{4.0(3.6)} & \mc{2}{c|}{4.2(3.8)} & \mc{2}{c}{4.2(3.9)}   \\\hline
\end{tabular}
\begin{tablenotes}
\footnotesize
\item[*] Additional uncertainties due to the angular distributions are 6.9\%, 8.5\%, 8.5\%, and 9.2\% for the 
$\dsp\dsa$, $\dsp\dsb$, $\dstp\dsa$, and $\dstp\dsb$, respectively.
\end{tablenotes}
\end{small}
\label{tab_sys}
\end{center}
\end{sidewaystable}

\section{Summary}

In summary, we observe the charmed strange meson pair $\dsstp\dsj$ production in $\ytwos$ decays and in $\EE$ 
annihilation at $\sqrt{s} = 10.52~\gev$ for the first time, where $\dsj$ is $\dsa$ or $\dsb$. We determine the 
products of branching fractions for the $\dsj$ production in $\ytwos$ decays to be 
\begin{eqnarray*}
\BR(\ytwos\to \dsp\dsa)\BR(\dsa\to \KM\dza ) & = & (1.6\pm 0.3\pm 0.2)\times 10^{-5}, \\ 
\BR(\ytwos\to \dstp\dsa)\BR(\dsa\to \KM\dza ) & = & (1.4\pm 0.4\pm 0.2)\times 10^{-5}, \\
\BR(\ytwos\to \dsp\dsb)\BR(\dsb\to \KM\dz ) & =  & (1.4\pm 0.4\pm 0.2)\times 10^{-5}, \\
\BR(\ytwos\to \dstp\dsb)\BR(\dsb\to \KM\dz ) & = & (0.9\pm 0.5\pm 0.1)\times 10^{-5}, 
\end{eqnarray*}
and
\begin{eqnarray*}
\BR(\ytwos\to \dsp\dsa)\BR(\dsa\to \ks\dzb ) & = & (0.84\pm 0.18\pm 0.14)\times 10^{-5}, \\ 
\BR(\ytwos\to \dstp\dsa)\BR(\dsa\to \ks\dzb ) & = & (0.82\pm 0.25\pm 0.18)\times 10^{-5}, \\
\BR(\ytwos\to \dsp\dsb)\BR(\dsb\to \ks\dm ) & =  & (0.69\pm 0.20\pm 0.21)\times 10^{-5}, \\
\BR(\ytwos\to \dstp\dsb)\BR(\dsb\to \ks\dm ) & = & (0.54\pm 0.31\pm 0.39)\times 10^{-5}. 
\end{eqnarray*}
We also determine the upper limit $\BR^{\rm UL}(\ytwos\to \dstp\dsb)\BR(\dsb\to \KM\bar{D}^0) < 2.5 \times 10^{-5}$ 
and $\BR^{\rm UL}(\ytwos\to \dstp\dsb)\BR(\dsb\to \ks\dm) < 1.4 \times 10^{-5}$ at 90\% C.L. We determine Born cross 
sections for continuum productions of the $\dsj$ at $\sqrt{s} = 10.52~\gev$ to be
\begin{eqnarray*}
\sigma^{\rm Born}(\EE\to\dsp\dsa)\BR(\dsa\to \KM\dza)  & = & (67\pm 8\pm 6)~\fb, \\
\sigma^{\rm Born}(\EE\to\dstp\dsa)\BR(\dsa\to \KM\dza) & = & (84\pm 11\pm 11)~\fb, \\
\sigma^{\rm Born}(\EE\to\dsp\dsb)\BR(\dsb\to \KM\dz)  & =  & (56 \pm 9\pm 13)~\fb, \\
\sigma^{\rm Born}(\EE\to\dstp\dsb)\BR(\dsb\to \KM\dz) & =  & (106\pm 17\pm 12)~\fb,
\end{eqnarray*}
and
\begin{eqnarray*}
\sigma^{\rm Born}(\EE\to\dsp\dsa)\BR(\dsa\to \ks\dzb)  & = & (34\pm 5\pm 4)~\fb, \\
\sigma^{\rm Born}(\EE\to\dstp\dsa)\BR(\dsa\to \ks\dzb) & = & (41\pm 6\pm 6)~\fb, \\
\sigma^{\rm Born}(\EE\to\dsp\dsb)\BR(\dsb\to \ks\dm)  & =  & (27\pm 6\pm 5)~\fb, \\
\sigma^{\rm Born}(\EE\to\dstp\dsb)\BR(\dsb\to \ks\dm) & =  & (51\pm 11\pm 9)~\fb,
\end{eqnarray*}

For comparison, $\sigma^{\rm Born}(\EE\to\MM) = 0.784~\nb$ at $\sqrt{s} = 10.52~\gev$ and $\BR(\ytwos\to\MM) = (1.93 
\pm 0.17)\%$~\cite{PDG}. We define the ratios $R_1 \equiv \BR(\ytwos\to \dsstp\dsj)/\BR(\ytwos\to\MM)$ for the 
$\ytwos$ decays and $R_2 \equiv \sigma^{\rm Born}(\EE\to\dsstp\dsj)/\sigma^{\rm Born}(\EE\to\MM)$ for the continuum 
productions. We obtain $R_1/R_2 = 9.7\pm 2.3 \pm 1.1$, $6.8\pm 2.1 \pm 0.8$, $10.2\pm 3.3 \pm 2.5$, and $3.4\pm 2.1 
\pm 0.5$ for the $\dsp\dsa$, $\dstp\dsa$, $\dsp\dsb$, and $\dstp\dsb$ final states in the $\dsj\to \KM\bar{D}^{(*)0}$ 
modes, respectively, where the uncertainty of $\BR(\ytwos\to\MM)$ is taken into account of the systematic 
uncertainties. Therefore, the strong decay is expected to dominate in $\ytwos\to\dsstp\dsj$ processes.

Here, we also determine the ratios of branching fractions to be 
\begin{eqnarray*}
\BR(\dsa\to \ks \dzb)/\BR(\dsa\to \KM \dza) & = & 0.48\pm 0.07 \pm 0.02, \\ 
\BR(\dsb \to \ks \dm)/\BR(\dsb \to \KM \dz) & = & 0.49 \pm 0.10 \pm 0.02,
\end{eqnarray*}
which are in good agreement with the expected 0.498 and 0.497 from isospin symmetry considering the phase space, 
since with $\ks$ only half of the neutral kaons can be reconstructed. The first ratio has same precision in 
comparison to the world average value~\cite{PDG}, while the second ratio is the first measurement of this quantity.

\acknowledgments

This work, based on data collected using the Belle detector, which was operated until June 2010, was supported by the 
Ministry of Education, Culture, Sports, Science, and Technology (MEXT) of Japan, the Japan Society for the Promotion 
of Science (JSPS), and the Tau-Lepton Physics Research Center of Nagoya University; 
the Australian Research Council including grants
DP210101900, 
DP210102831, 
DE220100462, 
LE210100098, 
LE230100085; 
Austrian Federal Ministry of Education, Science and Research (FWF) and FWF Austrian Science Fund No.~P~31361-N36;
National Key R\&D Program of China under Contract No.~2022YFA1601903,
National Natural Science Foundation of China and research grants
No.~11575017,
No.~11761141009, 
No.~11705209, 
No.~11975076, 
No.~12135005, 
No.~12150004, 
No.~12161141008, 
and
No.~12175041, 
and Shandong Provincial Natural Science Foundation Project ZR2022JQ02;
the Ministry of Education, Youth and Sports of the Czech Republic under Contract No.~LTT17020; the Czech Science 
Foundation Grant No. 22-18469S; 
Horizon 2020 ERC Advanced Grant No.~884719 and ERC Starting Grant No.~947006 ``InterLeptons'' (European Union);
the Carl Zeiss Foundation, the Deutsche Forschungsgemeinschaft, the
Excellence Cluster Universe, and the VolkswagenStiftung;
the Department of Atomic Energy (Project Identification No. RTI 4002) and the Department of Science and Technology of 
India; the Istituto Nazionale di Fisica Nucleare of Italy; 
National Research Foundation (NRF) of Korea Grant
Nos.~2016R1\-D1A1B\-02012900, 2018R1\-A2B\-3003643,
2018R1\-A6A1A\-06024970, RS\-2022\-00197659,
2019R1\-I1A3A\-01058933, 2021R1\-A6A1A\-03043957,
2021R1\-F1A\-1060423, 2021R1\-F1A\-1064008, 2022R1\-A2C\-1003993;
Radiation Science Research Institute, Foreign Large-size Research Facility Application Supporting project, the Global 
Science Experimental Data Hub Center of the Korea Institute of Science and Technology Information and 
KREONET/GLORIAD; the Polish Ministry of Science and Higher Education and the National Science Center;
the Ministry of Science and Higher Education of the Russian Federation, Agreement 14.W03.31.0026, 
and the HSE University Basic Research Program, Moscow; 
University of Tabuk research grants S-1440-0321, S-0256-1438, and S-0280-1439 (Saudi Arabia);
the Slovenian Research Agency Grant Nos. J1-9124 and P1-0135;
Ikerbasque, Basque Foundation for Science, Spain;
the Swiss National Science Foundation; 
the Ministry of Education and the Ministry of Science and Technology of Taiwan;
and the United States Department of Energy and the National Science Foundation.
These acknowledgements are not to be interpreted as an endorsement of any
statement made by any of our institutes, funding agencies, governments, or
their representatives. 
We thank the KEKB group for the excellent operation of the accelerator; the KEK cryogenics group for the efficient
operation of the solenoid; and the KEK computer group and the Pacific Northwest National Laboratory (PNNL) 
Environmental Molecular Sciences Laboratory (EMSL) computing group for strong computing support; and the National
Institute of Informatics, and Science Information NETwork 6 (SINET6) for valuable network support.

 \end{document}